\documentclass[aps,pra,twocolumn,
groupedaddress,10pt]{revtex4}
\usepackage{amsmath}
\usepackage{amssymb}
\usepackage{graphicx}
\usepackage{epsfig}
\usepackage{dcolumn}
\usepackage{bm}
\usepackage{bm}
\usepackage{blindtext}
\usepackage{natbib}
\usepackage{booktabs}

\def\be{\begin{equation}} \def\ee{\end{equation}}
\def\bea{\begin{eqnarray}} \def\eea{\end{eqnarray}}

\begin{document}

\title{A proposal for direct measurement on the quantum geometric potential}
\author{Jianda Wu}

\affiliation{Max-Planck-Institut f\"ur Physik komplexer Systeme, Dresden, 01187, Germany}

\begin{abstract}
The quantum geometric potential is a gauge invariant carrying
novel geometric features between any two energy levels or bands in quantum systems.
In generic time-dependent systems it gives a vital physical modification for
the instantaneous energy gaps, laying down more appropriate quantum adiabatic
conditions for both non-degenerate and degenerate systems.
Remarkably, for generic parameterized quantum systems,
the integration of the quantum geometric potential on a closed loop leads to a
novel type of quantized winding number, which is a quantum counterpart of the
Gauss-Bonnet theorem. The effects of the quantum geometric potential
had been indirectly supported by the experiments on the quantum adiabatic evolution,
however, a direct experimental observation
so far is lacking. In this paper we propose an interference measurement to
directly probe the quantum geometric potential, where the relevant parameters are
easily accessible by current
experimental apparatus. A direct confirmation of this new physical quantity could
motive further theoretical and experimental investigations as well as its potential
 real applications.
\end{abstract}

\date{\today}

\maketitle
\paragraph*{Introduction.---}
The investigation on the time-dependent systems almost started
simultaneously during the age when quantum mechanics is developing. The early study
of quantum adiabatic evolution in the time-dependent systems
yielded many important physical results, such as quantum adiabatic theorem
\cite{Born1928,Schwinger1937,Kato1950},
Landau-Zener transition \cite{Landau1932,Zener1932},
Gell-Mann-Low theorem \cite{Gellmann1951}. It later on
also led to the revelations of Berry phase \cite{berry1984proc},
and holonomy \cite{simon1983holonomy}, manifestly demonstrating the
beautiful geometric connections of the quantum wavefunctions.
Those progresses made in understanding the quantum adiabatic evolution
have led to a great deal of applications in quantum control,
quantum annealing, and quantum computation \cite{Oreg1984,Schiemann1993,Pillet1993,Jones2000,
farhi2001quantum,Childs2001,Zheng2005,ashhab2006decoherence,
das2008colloquium,Lidar2009,bason2012high,georgescu2014quantum,Lidar2016,Santos2017}.

The vital findings in the time-dependent systems
or more generic parameterized quantum systems, are not bounded in the particular
field of quantum adiabatic evolution \cite{Xiao2010, eckardt2017colloquium}.
Specifically, the Berry phase has been applied to
condensed matter systems, uncovering a diverse novel
phenomena, such as quantum charge pumping \cite{niu1990towards,PhysRevB.27.6083},
quantum spin Hall effect \cite{murakami2003,murakami20042,guo2008intrinsic},
and quantum anomalous Hall effect \cite{PhysRevLett.61.2015} . Furthermore,
the intensive researches on the Floquet periodic time-dependent systems in the past decade
have rolled out many exciting new areas, such as the fabrication of
artificial magnetic fields \cite{Aidelsburger2011}, dynamic quantum phase transitions \cite{Heyl2015},
Floquet topological band structure \cite{Jotzu2014}, etc..

Starting with the Berry connection, a berry phase is obtained
when an integral over the Berry curvature is carried out on a closed area in the
corresponding parameterized manifold. And the quantized first Chern
number is produced when the integration is extended to the whole
parameter space in the non-degenerate systems.
The non-Abelian Berry phase, a generalization of the original
Abelian one \cite{berry1984proc}, is later further introduced by
Wilczek and Zee \cite{wilczek1984appearance}.  It appears
in the quantum degenerate system with a $U(N)$ gauge field,
which is connected to the nontrivial topology, such as
the second Chern number and the Wilson loop \cite{PhysRevD.10.2445}.

For obtaining both of the Abelian and non-Abelian Berry phases,
an adiabatic evolution of the system is further required,
such that there is almost no transition
between instantaneous eigenstates with different instantaneous eigenvalues.
The Berry connection is a projection of the time-derivative of one instantaneous
eigen wavefunction to the one with the same instantaneous
eigenvalue. Then, because of the adiabatic evolution, the time-derivative one is almost
parallel to its corresponding instantaneous eigen wavefunction.
As a result, the Berry connection only
gives ``diagonal" connection information of the system.

The ``off-diagonal"
(inter-level) connection, namely, the projection of the time-derivative of
one instantaneous eigen wavefunction to the one with different instantaneous eigenvalue
was largely neglected. An inter-level Abelian gauge invariant in the non-degenerate
systems, referred as
quantum geometric potential (QGP), was first introduced in Ref.~\citep{Jianda2008},
which uncovers the telltale geometric features in the ``off-diagonal"
(inter-level) connections. The QGP appears in generic time-dependent systems
or more general parameterized systems. For the time-dependent systems,
it gives rise to
a better adiabatic condition to justify the quantum adiabatic evolution~\citep{Jianda2008}, whose
effects had been indirectly supported by an experiment on
the quantum adiabatic evolution \cite{Du2008}.
It is further found that the QGP can be straightforwardly generated
to the degenerate systems with non-Abelian gauge invariant.
For general parameterized systems,
the integration of the QGP on a close loops gives rise to
a novel type of quantized character, which is a quantum analogue of the Gauss-Bonnet
theorem \cite{Xu2017}. Based on all progresses being made,
it is compelling that
if one can directly measure the QGP. In this paper,
a simple interference measurement is proposed to
directly probe the quantum geometric potential.
A successful measurement on this quantity could
also stimulate further theoretical and experimental studies
along with its potential practical applications.

\paragraph*{The quantum geometric potential. ---}
This section is a brief retrospect of the QGP
in non-degenerate quantum systems \cite{Jianda2008,Zhang2010}.
Though the QGP can exist in generic parameterized systems,
it is more convenient to start from a parameterized non-degenerate
time-dependent $N$-level
Hamiltonian $\hat{H}(\textbf{x}(t))$ driven by $l$ real parameters
$\textbf{x}(t) = \{x_1(t), x_2(t),\cdots,x_l(t) \}$
as a function of time $t$.
Following the instantaneous eigen equations
\be
\hat{H}(\textbf{x})|\varphi_m(\textbf{x})\rangle =
e_m(\textbf{x})|\varphi_m(\textbf{x})\rangle,\; (m=1,2,\cdots,N),
\ee
in principal, a set of orthogonal eigenfunctions
$|\varphi_m(\textbf{x})\rangle$ with the corresponding
eigenvalues $e_m(\textbf{x})$ can be obtained at every instant moment $t$.
The ``diagonal" Berry connection for each instantaneous eigen function is defined as
\be
\mathcal{A}_{m}^{\mu}=i\langle\phi_m(\textbf{x})
|\partial_{x_\mu}|\varphi_m(\textbf{x})\rangle,\; (\mu = 1, 2, \cdots, l).
\ee
Then the quantum geometric potential for the non-degenerate systems
follows by
\be
\Delta_{mn} = \mathcal{A}_{n} - \mathcal{A}_{m} + \frac{d}{dt}\arg \langle\varphi_m|\dot{\varphi}_n\rangle,
\label{eq:ndqgp}
\ee
with the ``$\cdot$" denoting the derivative with respect to time.
In addition, $\mathcal{A}_{n(m)} \equiv \mathcal{A}_{n(m)}^{\mu}
\dot{x}_{\mu} \equiv i \langle\varphi_{n(m)}(\tau)|\dot{\varphi}_{n(m)}(\tau)\rangle$.
The QGP is gauge invariant under an arbitrary local $U(1)\otimes U(1)$ gauge transformation
$\left| {\varphi _{n(m)} (\tau )} \right\rangle  \to e^{if_{n(m)} (\tau )}
\left| {\varphi _{n(m)} (\tau )} \right\rangle \;\;\left[ {f_{n(m)} (0) = 0} \right]$
up to an initial constant phase where we conveniently choose it as zero.
Here, $f_{m(n)}(t)$ are smooth scalar functions. Starting from following
$U(1)$ local-gauge-invariant instantaneous eigen basis ($\hbar = 1$),
which is also the quantum adiabatic solutions from the time-dependent Schr\"odinger equation,
\be
\left| {\Phi _n^{adia} (\tau )} \right\rangle  = \exp \left\{ { - i\int_0^\tau  {\left[ {e_n (\lambda ) - \mathcal{A}_{n} (\lambda )} \right]d\lambda } } \right\}\left| {\varphi _n (\tau )} \right\rangle, \label{U1basis}
\ee
then $\Delta_{mn}$ can be formulated into following compact form,
\be
\Delta_{mn} = \frac{d}{dt}\arg \langle\tilde{\varphi}_m|\dot{\tilde{\varphi}}_n\rangle,
\label{QGP2}
\ee
where $|{\tilde{\varphi}}_n\rangle = \exp\{\int_0^\tau i \mathcal{A}_{n}(\lambda) d \lambda \} |\varphi_n\rangle$
is also local $U(1)$ gauge invariant. Eq.~(\ref{QGP2}) shows that
the ``off-diagonal" connection (written in the $U(1)$ invariant basis)
indeed contains the structure of the quantum geometric potential.
This becomes clearer when considering a spin-$\frac{1}{2}$
system processing in an external time-dependent
magnetic field, $\Delta_{mn}$
is related to the geodesic curvature of the path of the wavefunction
on the Bloch sphere, implying its direct geometric connection.
When applying $\Delta_{mn}$ to the time-dependent system,
an improved quantum adiabatic condition for the non-degenerate system can be established
for $n\neq m$ \cite{Jianda2008,Zhang2010,Zhao2011},
\be
\frac{{\left| {\langle\phi_m|\dot{\phi}_n\rangle } \right|}}
{{\left| {e_m (t) - e_n (t) + \Delta _{mn} (t)} \right|}}
\ll 1 \label{eq:U1adia},
\ee
which indicates ${e_m (t) - e_n (t) + \Delta_{mn} (t)} $
is more appropriate to describe the instantaneous energy gaps, whose
effects had been indirectly confirmed by the fidelity experiment
on the quantum adiabatic evolution \cite{Du2008}.

Remarkably, Ref.~\cite{Xu2017} demonstrated
a novel quantized character extracted out from the integration
of the quantum geometric potential over a closed area. The statement is summarized
as follows,
\be
2\pi \Theta  = \int_\mathcal{M} \mathcal{F}_{mn}  - \int_{\partial \mathcal{M}} {\Delta_{mn} dt}
\label{qgb}
\ee
where $\mathcal{M}$ denotes the integration area, while ${\partial \mathcal{M}}$ denotes
the boundary of $\mathcal{M}$. In addition
$\mathcal{F}_{mn} = F_n -F _m $,
where $F_{m(n)} = \frac{1}{2}F_{m(n)}^{\mu\nu} d\lambda^{\mu}\wedge d\lambda^{\nu}$
with
\be
F^{\mu\nu}_{m(n)} = \partial^\mu
\mathcal{A}^{\nu}_{m(n)} - \partial^\nu \mathcal{A}^{\mu}_{m(n)} \label{berrycurvature1}
\ee
being the
corresponding Berry curvature (gauge field) of $m(n)^{th}$ eigenvalue. It is shown that
the $\Theta$ in Eq.~(\ref{qgb}) is a local gauge invariant quantized character.
In a sharp contrast the Berry phase is only local gauge invariant up to a 2$\pi$ module.
The formal form of Eq.~(\ref{qgb}) bears similar structure of the
famous Gauss-Bonnet theorem, as such, it is referred as
the quantum counterpart of the Gauss-Bonnet theorem.
Furthermore, for the degenerate systems, after
properly pulling out the phase of the corresponding time evolution operator,
a non-Abelian-invariant QGP naturally emerges.
The non-Abelian-invariant QGP also plays an important role for better understanding
the quantum adiabatic evolution in the degenerate situation \cite{Xu2017}.

\paragraph*{An experimental proposal. ---}
The QGP is a general quantity in any generic parameterized system,
but a direct measurement is more accessible following the quantum adiabatic evolution
in time-dependent systems.
Here we propose a
neutron interferometric experiment to explicitly probe the QGP.
The proposal here is not limited to neutron, which can also
be conveniently carried out by other experimental platforms,
such as quantum optics, electrons, etc..

\begin{figure*}[t]
\begin{center}
\includegraphics[width=0.7\textwidth]{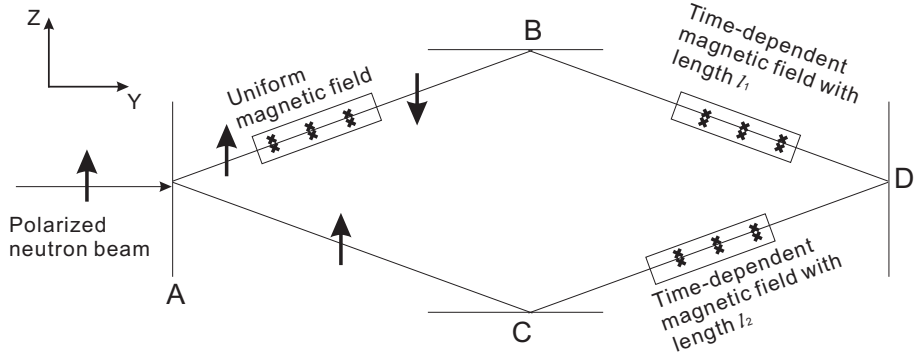}
\end{center}
\caption{Schematic experimental setup for measuring the QGP.}
\label{fig:experiment}
\end{figure*}
The proposed experimental setup is shown in Fig.~(\ref{fig:experiment}).
A polarized-spin-up($\left| { + z} \right\rangle$) neutron beam splits
into two beams at the position $(y=0,z=0)$ of interface $A$  due to Laue scattering.
Before the upper beam is reflected, a uniform magnetic field
is placed in the neutron propagation path
of upper beam, which rotates
the spin up to spin down($\left| { - z} \right\rangle$).
Two time-dependent magnetic fields are applied to the system after
the two beams are reflected. The beginning locations of the two fields
are both at $(y_0, \pm z_0)$, and the two fields have the same
time-dependence.
Since $|{\bf p}|$ (${\bf p}$ is the momentum of the neutron beams)
is the same between the upper and lower beams during
the propagation, thus two beams will arrive at the
time-dependent magnetic fields at the same time $t_0$, which should
be able to precisely controlled in the experiment.

At $t_0$ we turn on
the time-dependent magnetic field with the initial condition that
$\left| { + z} \right\rangle$ and $\left| { - z} \right\rangle$ as
the eigenstates of the Hamiltonian. The lengths of time-dependent
magnetic fields in upper and lower paths are $l_1$ and $l_2$.
For measuring the quantum geometric potential, the two lengths are different in general,
without loss of generality we set $l_2>l_1$, then
the upper and lower-branch of neutrons
will experience different evolution time with the same time-dependent magnetic fields.
The sufficient quantum adiabatic condition (similar as
the qualitative condition of Eq.~(\ref{eq:U1adia}))
derived in Ref.\cite{Jianda2008} needs to be imposed for
guaranteeing the time evolutions of the two neutron beams to be quantum adiabatic.
Finally the two neutron beams meet at destination $D$.
By measuring neutron spin coherent information at location $D$,
we can get both fo the information of the QGP and the $U(1)$-invariant basis (Eq.~(\ref{U1basis})).

Suppose now everything is under well control, then after the neutrons fly through the
time-dependent magnetic field in the upper path, the final state of neutron
becomes (assuming taking $t_1$ time for passing through that region)
\be
\left| {\Phi _ - ^{adia} (t_1 )} \right\rangle  = \exp \left\{ { - i\int_0^{t_1 } {\left[ {e_ -  (\lambda ) - \mathcal{A}_-(\lambda )} \right]d\lambda } } \right\}\left| {\varphi _ -  (t_1 )} \right\rangle
\ee
For the other path the final state becomes,
\be
\left| {\Phi _ + ^{adia} (t_2 )} \right\rangle  = \exp \left\{ { - i\int_0^{t_2 } {\left[ {e_ +  (\lambda ) - \mathcal{A}_+(\lambda )} \right]d\lambda } } \right\}\left| {\varphi _ +  (t_2 )} \right\rangle
\ee
As a result, their coherent intensity at point $D$ follows by
\begin{widetext}
\bea
&& I(t_1 ,t_2 ) =\frac{1}{2} \left| {\left| {\Phi _ - ^{adia} (t_1 )} \right\rangle  + \left| {\Phi _ + ^{adia} (t_2 )} \right\rangle } \right|^2
  = 1 +\frac{1}{2} \left\langle {{\Phi _ - ^{adia} (t_1 )}}
 \mathrel{\left | {\vphantom {{\Phi _ - ^{adia} (t_1 )} {\Phi _ + ^{adia} (t_2 )}}}
 \right. \kern-\nulldelimiterspace}{{\Phi _ + ^{adia} (t_2 )}} \right\rangle  +
 \frac{1}{2} \left\langle {{\Phi _ + ^{adia} (t_2 )}}
 \mathrel{\left | {\vphantom {{\Phi _ + ^{adia} (t_2 )} {\Phi _ - ^{adia} (t_1 )}}}
 \right. \kern-\nulldelimiterspace}
 {{\Phi _ - ^{adia} (t_1 )}} \right\rangle  \nonumber \\
  &=& 1 + \frac{1}{2} \left( {\exp \left\{ {i\int_0^{t_1 } {\left[ {e_ -  (\lambda ) - \mathcal{A}_- (\lambda )} \right]d\lambda } } \right\}\exp \left\{ { - i\int_0^{t_2 } {\left[ {e_ +  (\lambda ) - \mathcal{A}_+ (\lambda )} \right]d\lambda } } \right\}\left\langle {{\varphi _ -  (t_1 )}}
 \mathrel{\left | {\vphantom {{\varphi _ -  (t_1 )} {\varphi _ +  (t_2 )}}}
 \right. \kern-\nulldelimiterspace}
 {{\varphi _ +  (t_2 )}} \right\rangle  + c.c} \right) \nonumber\\
  &=& 1 + \frac{1}{2} \left( {\exp \left\{ {i\int_0^{t_1 } {e_ -  (\lambda )d\lambda }  - i\int_0^{t_2 } {e_ +  (\lambda )d\lambda } } \right\} } \right. \left. {\exp \left\{ { - i\int_0^{t_1 } {\mathcal{A}_- (\lambda )d\lambda }  + i\int_0^{t_2 } {\mathcal{A}_+ (\lambda )d\lambda } } \right\}\left\langle {{\varphi _ -  (t_1 )}}
 \mathrel{\left | {\vphantom {{\varphi _ -  (t_1 )} {\varphi _ +  (t_2 )}}}
 \right. \kern-\nulldelimiterspace}
 {{\varphi _ +  (t_2 )}} \right\rangle  + c.c} \right), \nonumber \\ \label{coherence1}
 \eea
 \end{widetext}
where $c.c$ is used to denote the conjugate term.
Now fine tuning $l_2$ for letting the length of $l_2$ just slightly
larger than the $l_1$, i.e., $l_2 = l_1 + \delta l$, then $t_2 = t_1 + \delta t$.
Substituting this back to the coherent intensity of Eq.~(\ref{coherence1}),
we obtain (noting the fact that $I(t_1,t_1)=1$),
\begin{widetext}
\bea
2(I(t_1 ,t_1  + \delta t) - I(t_1 ,t_1 ))  = \left( {\cos \left[ {\int_0^{t_1 } {\left[ {e_ -  (\lambda ) - e_ +  (\lambda ) + \Delta _{+-} (\lambda )} \right]d\lambda } } \right]} \right) \left| {\langle\varphi_-|\dot{\varphi}_+\rangle } \right| \delta t + O\left( {(\delta t)^2 } \right)
 \eea
$\Rightarrow$
\bea
\left. {\frac{{\partial I(t_1 ,t_2 )}}{{\partial t_2 }}} \right|_{t_2  \to t_1 }
 = \mathop {\lim }\limits_{\delta t \to 0} \frac{{I(t_1 ,t_1  + \delta t) - I(t_1 ,t_1 )}}{{\delta t}}
 =\frac{1}{2} \left| {\langle\varphi_-|\dot{\varphi}_+\rangle } \right| \cos \left[ {\int_0^{t_1 } {\left[ {e_ -  (\lambda ) - e_ +  (\lambda ) + \Delta _{+-} (\lambda )} \right]d\lambda } } \right]. \label{qgposcillation}
\eea
 \end{widetext}
From Eq.~(\ref{qgposcillation}),
as expected, the oscillation
of the coherent density is not longer solely determined by the
instantaneous energy gap but also modified by the QGP.
By measuring the differential coherent density, after subtracting out the phase
contribution from the dynamic part from the instantaneous energy gap,
we can directly verify the existence of the QGP ---
the quantum ``geodesic curvature" of any two sub-manifolds with two different instantaneous eigenstates
in time-dependent Hamiltonian. It is worth to note that
when the time-evolution completes a cycle in Eq.~(\ref{qgposcillation}),
the proposal can also help probe the influence of the
quantized character originating from the QGP [Eq.~(\ref{qgb})].
The spirit of the proposal in neutron experiment
can be equivalently converted to other type of experimental platforms, as well as
applied to generic parameterized quantum systems.
Now let's move to an explicit model which can be realized by real experiments.

\paragraph*{An explicit experimental accessible model. ---}
Consider following spin-$\frac{1}{2}$ particle in a rotating magnetic field. The Hamiltonian of the system is
\bea
h(\tau ) = \eta \sigma _z  + \xi \left[ {\sigma _x \cos (2K\eta \tau ) + \sigma _y \sin (2K\eta \tau )} \right]
\eea
with time-independent eigenvalues $E_\pm   = \pm \sqrt {\eta ^2  + \xi ^2 }$.
Properly choosing phases, two adiabatic orbits (instantaneous eigenstates)
can be expressed as
\bea
&&\left| {\varphi _ +  (\tau )} \right\rangle  = \cos (\theta /2)\left| 0 \right\rangle  + e^{2iK\eta \tau } \sin (\theta /2)\left| 1 \right\rangle  \\
&& \left| {\varphi _ -  (\tau )} \right\rangle  = \sin (\theta /2)\left| 0 \right\rangle  - e^{2iK\eta \tau } \cos (\theta /2)\left| 1 \right\rangle
 \eea
where $\cos \theta = \eta/\sqrt{\eta^2+\xi^2}$. It is straightforward to obtain
$\Delta_{+-} = 2 K \eta \cos \theta$ and
$| {\langle\varphi_-|\dot{\varphi}_+\rangle } | = | K \eta \sin \theta |$
$= |K \eta \xi /\sqrt{\eta^2 + \xi^2}|$, both are time-independent in this simple model.

Now let's set parameters on neutron, $\eta \sim$ 500 Gauss$\times$$\mu_n$
(${\rm{neutron\;magneton}}\;\mu_n = -1.913\times {\rm nuclear\;magneton} = 5.05
\times 10^{-27} J/{\rm Tesla} = 9.66\times 10^{-27}J/{\rm Tesla}$) and $\xi \sim 5$
Gauss$\times$$\mu_n$,
whose corresponding energy scales are $\eta\sim 4.78\times10^{-28} J \sim 721$KHz
and $\xi \sim 7.21$KHz. Then if we choose $K=5$, then the rotation frequency
on the $x$-$y$ plane is around $7.21$MHz which is doable under current experimental
conditions. Furthermore we have $\cos\theta \approx 0.99995$ which indicates
$\theta$ is almost zero which are well consistent with initial condition.
With those chosen parameters, it is easy to verify that
$| {\langle\varphi_-|\dot{\varphi}_+\rangle } |/|E_- - E_+ - \Delta_{+-}|
=  K/(2(\xi/\eta + \eta/\xi)+2K\eta/\xi) < 5/1000 \ll 1$ which satisfies
the sufficient adiabatic condition derived in Ref.~\cite{Jianda2008} with
the fidelity $F>99/100$. This indicates
the system's time evolution can be well described
by the quantum adiabatic evolution. Since
$\Delta_{+-} = 2K\eta \cos \theta \approx 10\eta $ and $E_- - E_+ \approx - 2 \eta$,
so under this set of parameters one should observe the behavior of $dI/dt \propto \cos(8\eta \tau)$
with an oscillation frequency of around 5.77MHz, about four times of the
(instantaneous) energy gap. Thus with the chosen parameters
the QGP has much more contribution than that of the energy gap of the system,
which should be easily identified in the experiments.
If the $z$-direction coupling is removed in the above
Hamiltonian, then the effects of QGP disappears,
the system's dynamics will be governed
by the energy gap of the system.
Actually if the oscillation of magnetic
field along $z$-direction is further turned on,
one can have a better proposal, whose details will be deferred
to a future publication.


\paragraph*{Discussions and conclusions. ---}
At first glance one may think the QGP as a certain transforming form of
the Berry curvature\cite{Xiao2010}
since both of them are induced by the Berry connection in certain ways,
however, they are complementary.
In any instantaneous eigenstate, the $n^{th}$
instantaneous orbital in the Hilbert space $\left| {\varphi _n (\tau )} \right\rangle$
of a system, all of its parameters compose a sub-manifold (consider the
total Hamiltonian composed by all instantaneous eigenvalues and eigenstates
as a total manifold). Along its time evolution direction, the Berry curvature
(Eq.~(\ref{berrycurvature1}))
is defined in the corresponding sub-manifold classified by the quantum number $n$.
As a result, the Berry curvature
is locally defined and solely lives in a $n^{th}$ manifold.
The difference now becomes crystal clear, although QGP is
also a local-gauge invariant term,
but the more important fact is that the QGP relates any two
sub-manifolds characterized by two different quantum numbers during the time-dependent evolution.
The difference is manifest when considering a single spin-$\frac{1}{2}$
coupled to a general time-dependent magnetic field,
$H =  B {\bf{\sigma}}  \cdot { \bf{n}}(\tau )$
with the magnetic moment absorbed into the magnetic field $B$, and
the unit vector $\bf{n}(\tau )= (\sin \theta(\tau) \cos \phi(\tau), \sin \theta(\tau) \sin \phi(\tau), \cos \theta(\tau))$
denotes the direction of the field.
Its Berry curvature becomes $\sin \theta /2$
(corresponding to the sub-manifold with eigenvalue $B$),
while the corresponding QGP follows by,
\be
\Delta _{ +  - }  = \frac{{\dot \theta \ddot \phi \sin \theta  + 2\dot \theta ^2 \dot \phi \cos \theta  + \dot \phi ^3 \sin ^2 \theta \cos \theta  - \dot \phi \ddot \theta \sin \theta }}{{\dot \theta ^2  + (\dot \phi \sin \theta )^2 }},
\ee
which is equal to the geodesic curvature of the unit 2-sphere
multiplied with the 2-sphere measure distance
$(\dot \theta ^2  + (\dot \phi \sin \theta )^2 )^{1/2}$ \cite{Jianda2008}.

Once the QGP is restricted into only one sub-manifold, say $n=m$, the QGP
just disappears. In this sense the QGP is a physical quantity reflecting
important geometric characteristics between different sub-manifolds in
the generic parameterized quantum system.
The emergence of the Berry phase implies that one
can not fix the phase factor globally during the time evolution.
The QGP further indicates that even locally at any instant
moment the phase factor can not be fixed arbitrarily either. The two
quantities are complimentary between each other, laying down a better
and more complete picture for understanding the
geometric characteristics and even the topology in the
time-dependent and generic parameterized
quantum systems.

In summary, we propose a straightforward interference measurement
to directly probe the QGP, following which a measurement on the
influence of the
corresponding quantized character is also possible after a cyclic
time-evolution. It is expected that the proposal
can be carried out by certain experiments in the near future. 
The physical and
geometric implications of the QGP are also discussed and analysed
following the simple spin-$\frac{1}{2}$ model. Based on the unique
features of the Berry phase and the QGP, we conclude that the two
quantities are complementary.

\paragraph*{Acknowledgement. ---}
JW Thanks helpful discussion with Dr. Meisheng Zhao.

\bibliography{GINV}



\end{document}